\documentclass[aps,prl,floatfix,superscriptaddress,10pt,twocolumn]{revtex4}
\usepackage{color,amsmath,amssymb,graphicx,latexsym,subfigure}
\usepackage{threeparttable,multirow,txfonts,makecell,tabularx}
\DeclareMathAlphabet{\mathcal}{OMS}{cmsy}{m}{n}
\usepackage{appendix}
\usepackage{hyperref}
\newcommand{\nc}{\newcommand}
\nc{\beqa}{\begin{eqnarray}}  
\nc{\eeqa}{\end{eqnarray}}  

\begin{document}
	
	\title{ Constraints on new physics around the MeV scale with cosmological observations}

	\author{Shihao Deng}

		\affiliation{Department of Physics and Chongqing Key Laboratory for Strongly Coupled Physics, Chongqing University, Chongqing 401331, P. R. China}
	\author{Ligong Bian}
	\email{lgbycl@cqu.edu.cn}

	\affiliation{Department of Physics and Chongqing Key Laboratory for Strongly Coupled Physics, Chongqing University, Chongqing 401331, P. R. China}
\affiliation{ Center for High Energy Physics, Peking University, Beijing 100871, P. R. China}

	\begin{abstract}
	
We investigate the joint effect of the cosmological phase transitions, thermal light dark matter, and the lepton asymmetry on the big bang nucleosynthesis and cosmic microwave background. We find that all of them can modify the predictions of the effective number of neutrino species and primordial nucleosynthesis. In turn, we observe that: 1) the cosmological observations can exclude slow and strong phase transitions with strength even smaller than $\mathcal{O}(10^{-3}-10^{-2})$; 2) a much larger portion of dark matter mass region is excluded when the phase transition temperature is closer to 1 MeV; and 3) the magnitude of the non-vanishing neutrino lepton asymmetry is limited to be around $\mathcal{O}(10^{-2}-10^{-1})$ depending on the phase transition strength. These phase transitions can produce stochastic gravitational wave background to be probed by pulsar timing array experiments. 
	\end{abstract}
	
	\maketitle
\section{Introduction}
The cosmological first-order phase transitions (PTs) are predicted by many well-motivated new physics models~\cite{Losada:1996ju,Cline:1996mga,Laine:1996ms,Bodeker:1996pc}. 
The  first-order PTs are expected to produce stochastic gravitational wave background (SGWB)~\cite{Caprini:2015zlo,Caprini:2019egz}, 
explain the source of primordial magnetic fields~\cite{Vachaspati:1991nm,Di:2020ivg} and the origin of the baryon asymmetry of the Universe~\cite{Morrissey:2012db}. 
The SGWB produced by the first-order PTs is one of the main scientific goals of many gravitational detectors, such as LIGO~\cite{TheLIGOScientific:2014jea,Romero:2021kby}, LISA~\cite{Audley:2017drz}, $Taiji$~\cite{Guo:2018npi}, NANOGrav~\cite{NANOGrav:2021flc}, PPTA~\cite{Xue:2021gyq}, and SKA~\cite{Carilli:2004nx}.
Since both the types of electroweak PT and QCD PT
in the Standard Model of particle physics are {\it cross-over}~\cite{DOnofrio:2014rug,Fodor:2001pe}, observing the SGWB relics of the first-order PTs would help to probe the parameters of new physics beyond the Standard Model (BSM)~\cite{Caprini:2015zlo,Bian:2021ini,Caprini:2019egz,Cai:2017cbj,Caldwell:2022qsj}.

The low-scale first-order PTs can occur in thermal dark sectors~\cite{Breitbach:2018ddu}, and QCD when lepton asymmetry shows up~\cite{Schwarz:2009ii,Wygas:2018otj,Middeldorf-Wygas:2020glx,Gao:2021nwz}. 
 PTs in the dark sector are also of great interest since they have the chance to modify the dark matter predictions through change particles masses ~\cite{Cohen:2008nb,Baker:2016xzo,Croon:2020ntf,Hashino:2021dvx,Dimopoulos:1990ai,Elor:2021swj,Bian:2018mkl,Bian:2018bxr,Deng:2020dnf,Baker:2017zwx,Darme:2019wpd}, interactions~\cite{Boddy:2012xs}, or the dark matter production dynamic in the early universe~\cite{Chao:2020adk,Baker:2019ndr,Azatov:2021ifm,Hong:2020est}.
Recently, the EMPRESS survey~\cite{Matsumoto:2022tlr} reported a smaller primordial helium abundance in comparison with the prediction of the Standard Model~\cite{Pitrou:2018cgg}, which may suggest the existence of a nonzero lepton asymmetry~\cite{Escudero:2022okz,Burns:2022hkq} that can originate from 
Affleck-Dine baryogenesis~\cite{Affleck:1984fy,Casas:1997gx}, resonant leptogenesis~\cite{Pilaftsis:2003gt} and topological defects~\cite{Bajc:1997ky}, and may generate the observed baryon asymmetry~\cite{March-Russell:1999hpw,Borah:2022uos}. 

It's known that MeV-scale dark matter can affect the neutrino decoupling process and therefore the cosmic microwave background (CMB) and Big Bang nucleosynthesis (BBN), since they can change the evolution of the energy density in the Universe~\cite{Serpico:2004nm,Chu:2022xuh,Sabti:2019mhn,Escudero:2018mvt,Sabti:2021reh,EscuderoAbenza:2020cmq,Sabti:2019mhn}. 
The appearance of the lepton asymmetry in the neutrino sector could also alter the BBN predictions since it can change the rate of proton-to-neutron conversions in the early Universe~\cite{Sarkar:1995dd,Serpico:2005bc,Chu:2006ua,Iocco:2008va,Simha:2008mt,Mangano:2011ip,Pitrou:2018cgg}. 
Ref~\cite{Bai:2021ibt} note that energy injection from MeV scale first-order PT can yield photon reheating and/or neutrino reheating and change the time-temperature relation, which consequently affects the effective number of neutrino species and the primordial abundance of helium and deuterium. More recently, Ref.~\cite{Liu:2022lvz,Liu:2021svg} shows that the PT's duration would further induce different energy density evolution history in the early Universe, therefore one can expect that the neutrino decoupling process and the BBN process would be further modified.

In this study, we perform a joint analysis on PTs, dark matter, and neutrino lepton asymmetry with cosmological observations of CMB and BBN. For the first time,
we place constraints on thermal light dark matter and the lepton asymmetry when the first-order PT's effect was taken into account. 
We observe that the appearance of a MeV-scale first-order PT would strengthen the constraints on dark matter mass and the magnitude of the neutrino lepton asymmetry.

 
\section{Thermal dynamics with PTs}   

Generally, light dark matter particles can interact with both neutrinos and electrons. The electrophilic scenario can be obtained with the assistance of lepton portal interaction~\cite{Bai:2014osa}, scalar portal interaction originating from dimension five operator~\cite{Chen:2018vkr}, or an extra $Z^\prime$~\cite{Bell:2014tta,Gu:2017gle,Chen:2021ifo}, while the neutrinophilic scenario can be realized when high dimensional operators~\cite{Kelly:2019wow,Kelly:2020pcy} or an extra U(1) gauge symmetry is introduced~\cite{Nomura:2017jxb,Nomura:2017wxf}. 
Without loss of generality and for simplicity, we consider PT's impact in purely electrophilic scenarios and purely neutrinophilic scenarios.
First-order PTs proceed through true vacuum bubbles nucleation and percolation with the nucleation rate~\cite{Coleman:1977py,Enqvist:1991xw}:
$\Gamma(t)=\Gamma_{0}e^{\beta t}\,,$ 
where the $\beta$ characterizes the PT rate or the true vacuum bubbles nucleations rate, and pre-factor can be estimated as $\Gamma_0^{1/4}=\left (4\pi^3g_\star/45  \right ) ^{1/2}(T_{\mathrm{p}}^2/m_\mathrm{Pl})e^{-\beta/8H_\star}$ in the radiation dominant universe with $T_{\mathrm{p}}$ being the PT temperature and $H_{*}$ being the Hubble parameter at the PT temperature, the Planck mass is $m_{\rm Pl} \approx 1.22\times 10^{19}\,\text{GeV}$~\cite{He:2022amv}.  At the PT time, the PT's inverse duration is $\beta/H_{*}\equiv \beta/H(t_{\mathrm{p}})$ and the PT's strength is $\alpha\equiv\Delta V/\rho_{\mathrm{r}}(t_{\mathrm{p}})$ where $\Delta V$ denotes the energy density difference between the false and true vacua. We consider the PT occurs as the averaged probability of the false vacuum $F(t_{\mathrm{p}})=0.7$. The $F(t)$ can be calculated through~\cite{Turner:1992tz}:
$F(t)=\exp \left[-(4 \pi/3) \int_{t_{i}}^{t} \mathrm{~d} t^{\prime} \Gamma\left(t^{\prime}\right) a^{3}\left(t^{\prime}\right) r^{3}\left(t, t^{\prime}\right)\right]\,
$, where $t_{i}$ is the time when PTs starts and $r(t,t')\equiv\int_{t'}^{t} a^{-1}(\tau) d\tau$ is the comoving radius of true vacuum bubbles. Before PTs, all the fields settle in the false vacuum with $F(t<t_{i})=1$.

 \begin{figure}[!htp]
\begin{center}
\includegraphics[width=0.4\textwidth]{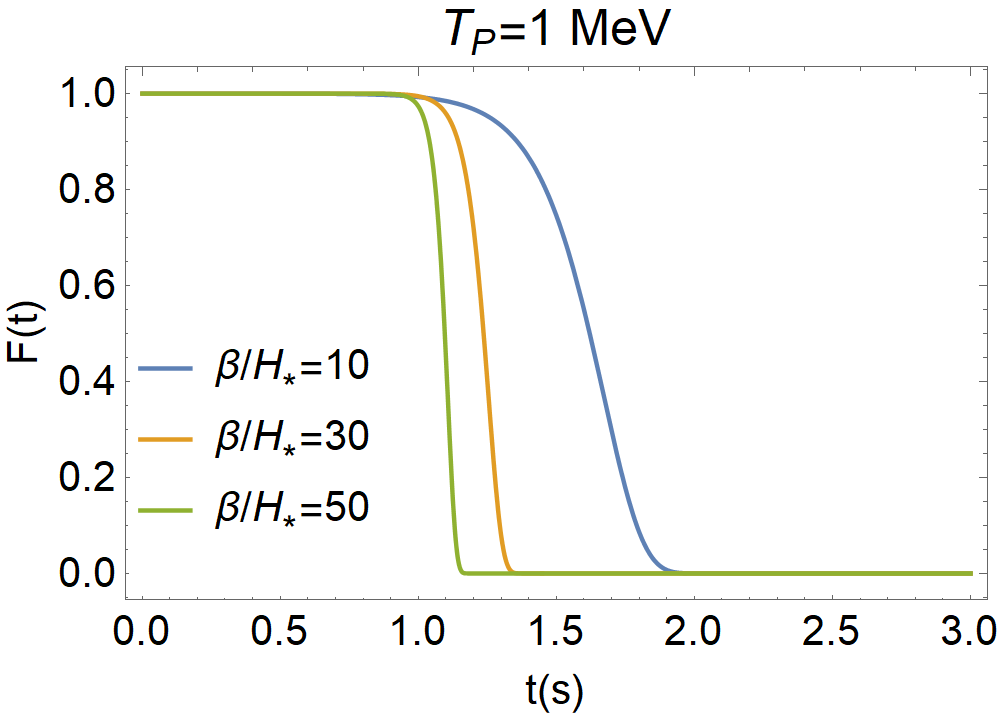}
\includegraphics[width=0.4\textwidth]{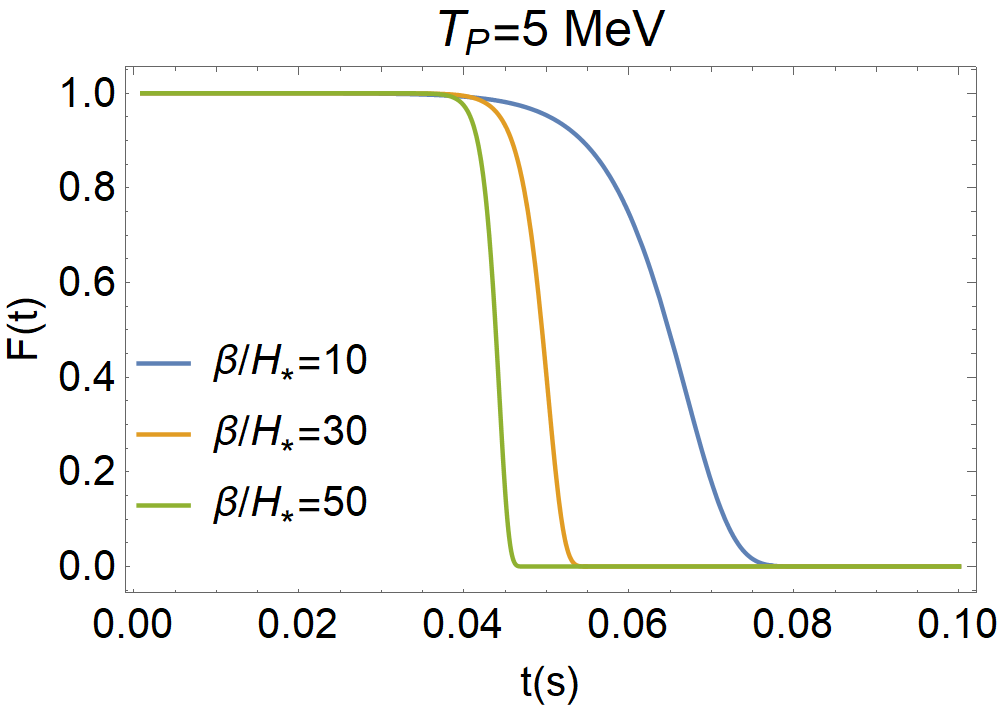}
		\caption{$F(t)$ as a function of t, for $\beta/H_{*}=10$, $30$, and $50$, respectively with $T_{\rm P}=1\,\text{MeV}$ (upper plot) and $T_{\rm P}=5\,\text{MeV}$ (lower plot).}
		\label{fig:Ft}
		\end{center}
	\end{figure}
	
As the PT proceeds, the false vacuum energy density $\rho_{\mathrm{vac}}(\equiv F(t)\Delta V)$ transfers into the background plasma and could yield photon (neutrino) reheating, causing an increase in their temperature and a subsequent decrease (increase) in $N_{\rm eff}= 3\times \left(11/4 \right)^{4/3} \left(T_\nu/T_\gamma\right)^4$. 
Around $1\,\text{MeV}$, the neutrinos decouple, so the injection of PT's energy around this time will have a significant effect on $N_{\rm eff}$.
 As illustrated in Fig.~\ref{fig:Ft}, the value of $F(t)$ changes much faster for a larger value of $\beta/H_{*}$ at the fixed PT temperatures.   $T_{\rm P}$ affects the period of the decrease of $F(t)$, as $T_{\rm P}$ increases, the magnitude of the $F(t)$ would decrease earlier.
 For $T_{\rm P}=1\,\text{MeV}$, $F(t)$ starts to fall just around $1{ \rm s}$, which would significantly affect the neutrino decoupling process. While $F(t)$ decreases to zero long before neutrino decoupling for $T_{\rm P}=5\,\text{MeV}$, and thus the effect on neutrino decoupling will be weaker, so $N_{\rm eff}$ changes even less compared to the Standard Model.


 We first study the early thermodynamics of the universe with MeV-scale thermally electrophilic dark sectors, where photon reheating driven by the first-order PTs occurs through energy injection. 
 More explicitly, we extend the corresponding temperature evolution equations given in Ref~\cite{Escudero:2018mvt} to include the PT's dynamics,  
\begin{equation}\label{eq:dTnudt_DM_e}
\begin{aligned} \frac{dT_{\gamma}}{dt} = &-   
 \Bigg( 4 H \rho_{\gamma} + 3 H \left( \rho_{e} + p_{e}\right) +  3 H \left( \rho_{\chi} + p_{\chi}\right) \\
 &+ 3 H \, T_\gamma \frac{dP_\text{int}}{dT_\gamma}+  \frac{\delta \rho_{\nu_e}}{\delta t} + 2 \frac{\delta \rho_{\nu_{\mu}}}{\delta t}+\frac{d\rho_{\mathrm{vac}}}{dt} \Bigg) \bigg/f(T_\gamma)\,,\\
\frac{dT_\nu}{dt} = &-( 12  H \rho_\nu  -   \frac{\delta \rho_{\nu_e}}{\delta t} -2 \frac{\delta \rho_{\nu_{\mu}}}{\delta t})\big/(3 \, \frac{\partial \rho_\nu}{\partial T_\nu })\,.
\end{aligned} 
\end{equation}
Here, $f(T_\gamma)=\frac{\partial \rho_{\gamma}}{\partial T_\gamma} + \frac{\partial \rho_e}{\partial T_\gamma} + \frac{\partial \rho_{\chi}}{\partial T_\gamma} +T_\gamma \frac{d^2 P_\text{int}}{dT_\gamma^2} $, and the energy exchange rates ${\delta \rho_{\nu_e}}/{\delta t}$ and ${\delta \rho_{\nu_\mu}}/{\delta t}$ are:
\begin{equation}\label{eq:energyrates_nu_SM}
\begin{aligned}
\left. \frac{\delta \rho_{\nu_e}}{\delta t} \right. &= \frac{G_F^2}{\pi^5}\left[\left( 1 + 4 s_W^2 + 8 s_W^4 \right) F(T_\gamma,T_{\nu_e}) + 2  F(T_{\nu_\mu},T_{\nu_e}) \right] \, ,\\
\left. \frac{\delta \rho_{\nu_\mu}}{\delta t} \right. &= \frac{G_F^2}{\pi^5}\left[\left( 1 - 4 s_W^2 + 8 s_W^4 \right)  F(T_\gamma,T_{\nu_\mu}) -   F(T_{\nu_\mu},T_{\nu_e}) \right] \, ,
\end{aligned}\nonumber
\end{equation}
with $F(T_1,T_2) = 32\, (T_1^9-T_2^9) + 56 \, T_1^4\,T_2^4 \, (T_1-T_2)\,$, and where $G_F= 1.1664\times10^{-5}\,\text{GeV}^{-2}$ is the Fermi constant, and $s_W^2 = 0.223$ accounts for the Weinberg angle~\cite{ParticleDataGroup:2018ovx}.
Finite temperature corrections are accounted for by $P_{\rm int}$ and its derivatives~\cite{Escudero:2018mvt}.
In the above equations, $\rho_i$ and $p_i$ correspond to the energy density and pressure of a given particle respectively, $H = \sqrt{(8\pi/3)\,({\sum_i \rho_i}+\rho_{\mathrm{vac}})/m_{\rm Pl}^2}$ is the Hubble parameter. 
 
Similarly, for the neutrino reheating case, we consider the neutrinophilic dark sectors, the corresponding temperature evolution equations read:
 \begin{align}\label{eq:dTnudt_DM_nu}
\frac{dT_\nu}{dt} &= -\frac{ 12 H \rho_\nu +3 H( \rho_{\chi} + p_{\chi})  - \frac{\delta \rho_{\nu_e}}{\delta t} - 2 \frac{\delta \rho_{\nu_\mu}}{\delta t} + 3\, \frac{d\rho_{\mathrm{vac}}}{dt} }{3 \, \frac{\partial \rho_\nu}{\partial T_\nu } +  \frac{\partial \rho_{\chi}}{\partial T_\nu } }\,,\nonumber\\
\frac{dT_{\gamma}}{dt}  &=- \frac{  4 H \rho_{\gamma} + 3 H \left( \rho_{e} + p_{e}\right) + 3 H \, T_\gamma \frac{dP_\text{int}}{dT_\gamma}+ \frac{\delta \rho_{\nu_e}}{\delta t} + 2 \frac{\delta \rho_{\nu_\mu}}{\delta t} }{ \frac{\partial \rho_{\gamma}}{\partial T_\gamma} + \frac{\partial \rho_e}{\partial T_\gamma} +T_\gamma \frac{d^2 P_\text{int}}{dT_\gamma^2} } \,.
\end{align}

  \begin{figure}[!htp]
\begin{center}
\includegraphics[width=0.4\textwidth]{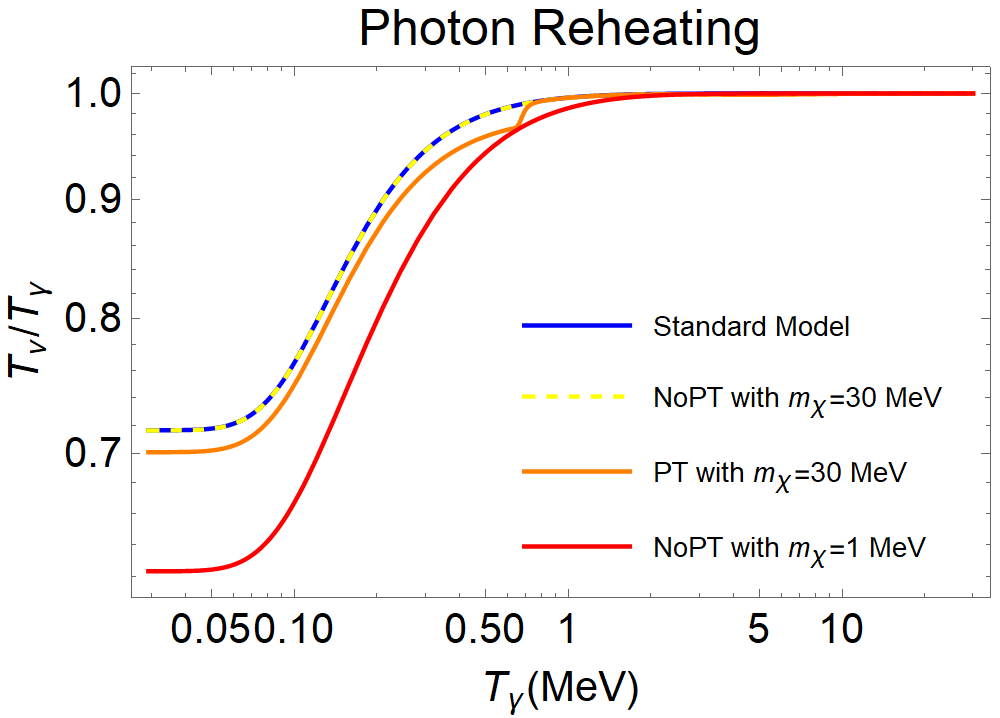}
\includegraphics[width=0.4\textwidth]{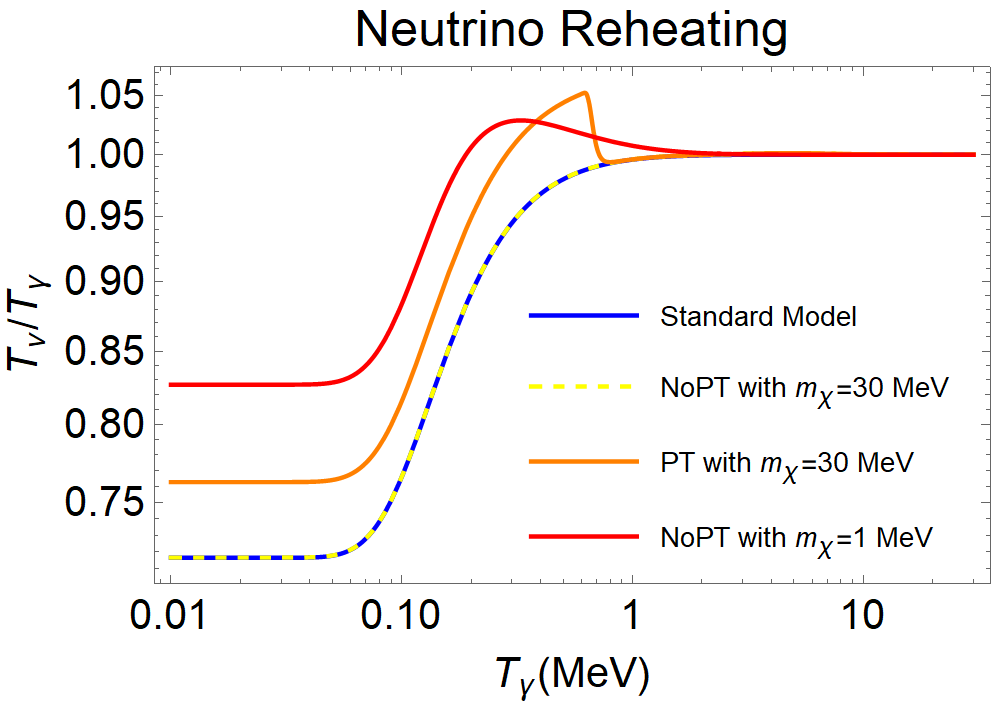}
		\caption{The top (bottom) plot shows the evolution of $T_\nu/T_\gamma$ with $T_\gamma$ for photon (neutrino) reheating, respectively. The dashed yellow lines correspond to the existence of a vector boson dark matter with $m_\chi=30\,\text{MeV}$ without PT, which excellently agrees with the Standard Model case (blue lines). The orange lines show the scenario where we consider a PT with the PT temperature $T_{\rm P} = 1\,\text{MeV}$, the PT's strength $\alpha = 0.01$, and the PT's inverse duration $\beta/H_{*} = 10$. The influence of the mass of dark matter on neutrino decoupling is shown with red curves, where we choose $m_\chi=1\,\text{MeV}$ for a significant comparison.}
		\label{fig:TnuT}
		\end{center}
	\end{figure}

We solve the time evolution equations for $T_\gamma$ and $T_\nu$ starting from $T_\gamma = T_\nu = 30\,\text{MeV}$, when the neutrino and electron are in thermal equilibrium. According to $t = 1/(2H)$, the starting time for the evolution is $t_0 \sim 7\times 10^{-4} \, \text{s}$, and we evolve the system until $t_{\rm final} = 5\times 10^4\,\text{s}$ where the electrons and positrons have already annihilated away. By solving this set of differential equations, we can find all the key background evolution quantities as a function of time, such as Hubble rate, temperature, etc. Technically,  we modify the publicly available versions of \texttt{nudec\_BSM}~\cite{Escudero:2018mvt,EscuderoAbenza:2020cmq}  to take into account the PT dynamics and use it to compute the background thermodynamics and $N_{\rm eff}$, which is crucial for CMB observations.

	     \begin{figure*}[!htp]
\begin{center}
\includegraphics[width=0.4\textwidth]{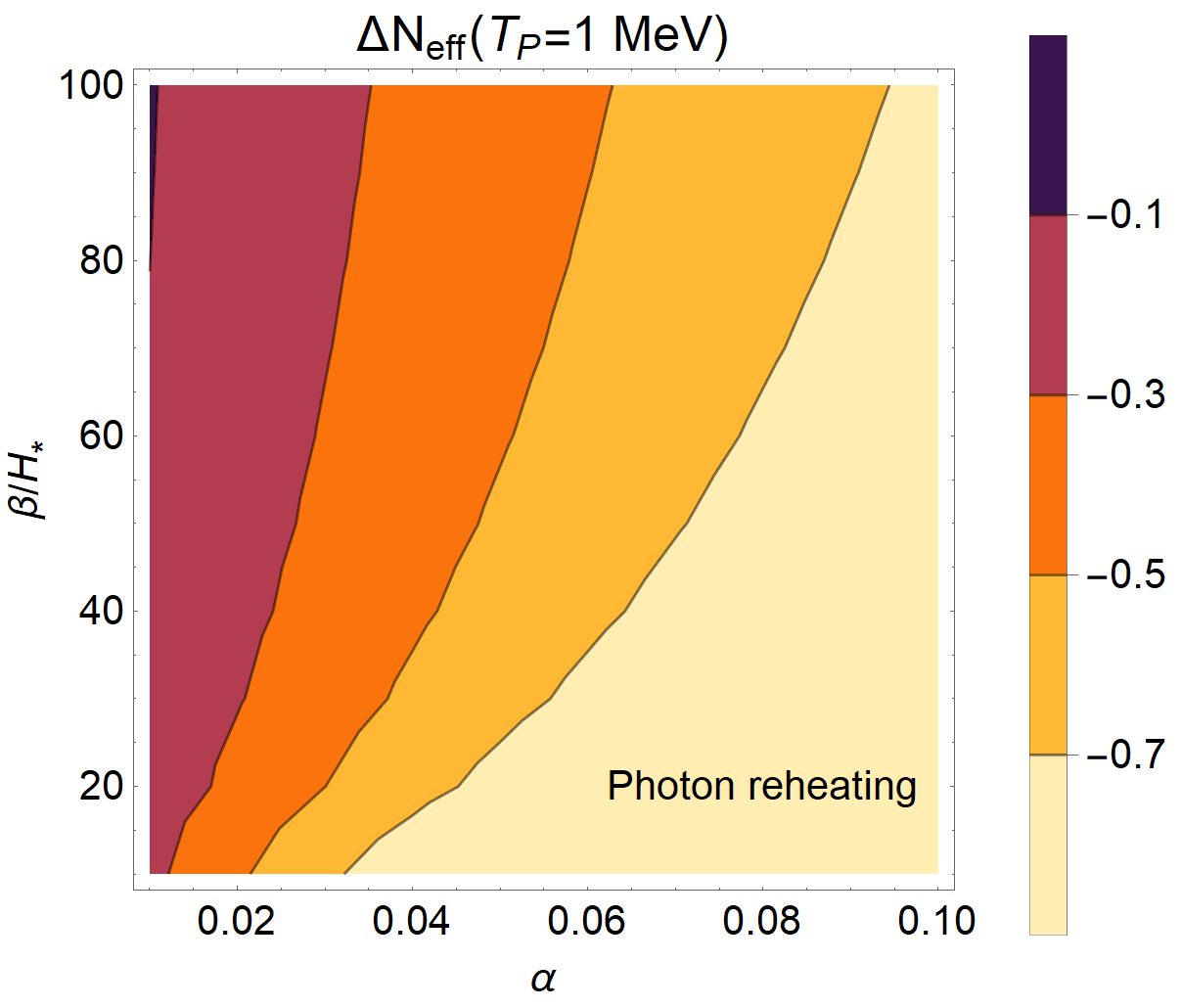}
\includegraphics[width=0.4\textwidth]{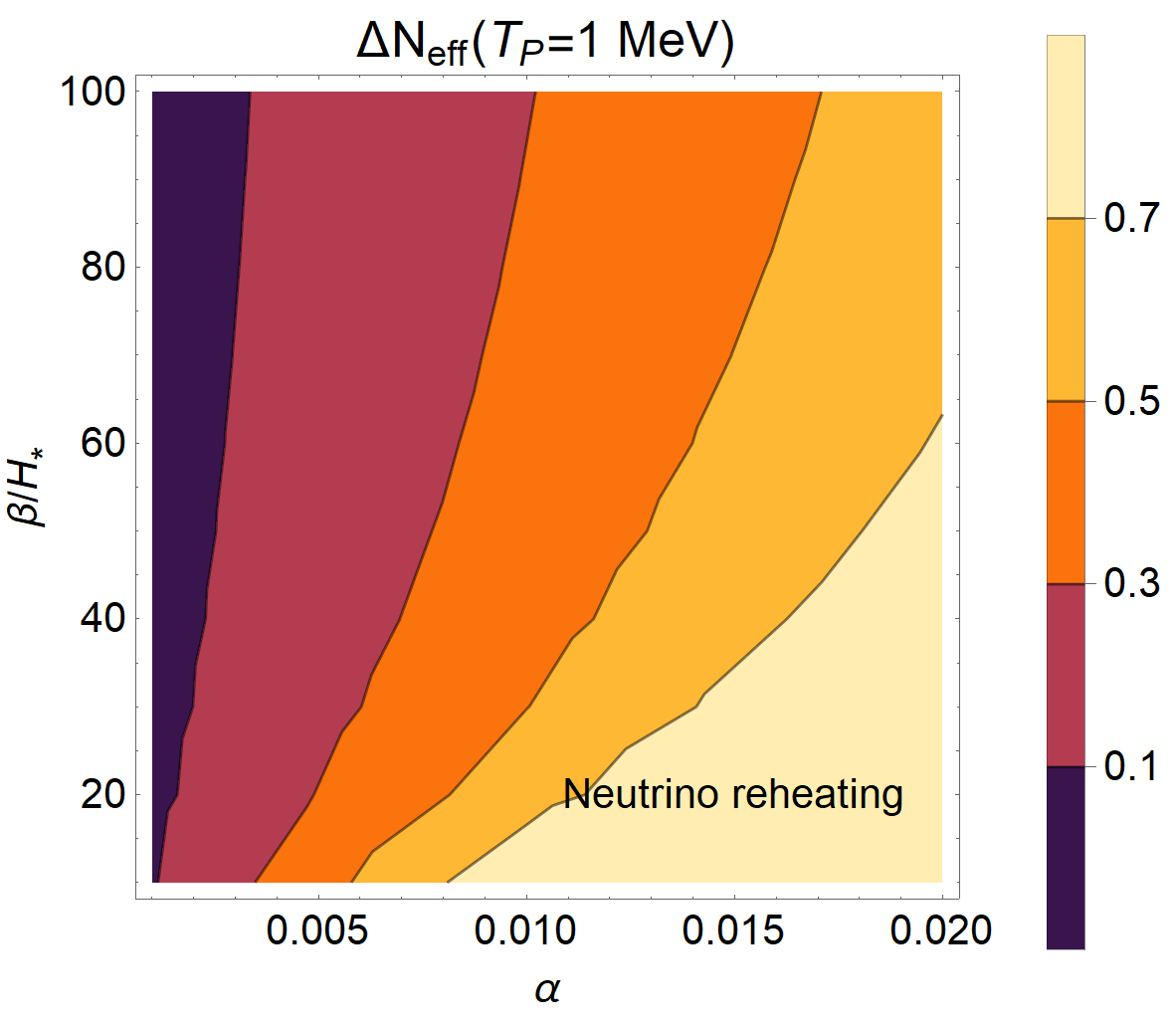}
\includegraphics[width=0.4\textwidth]{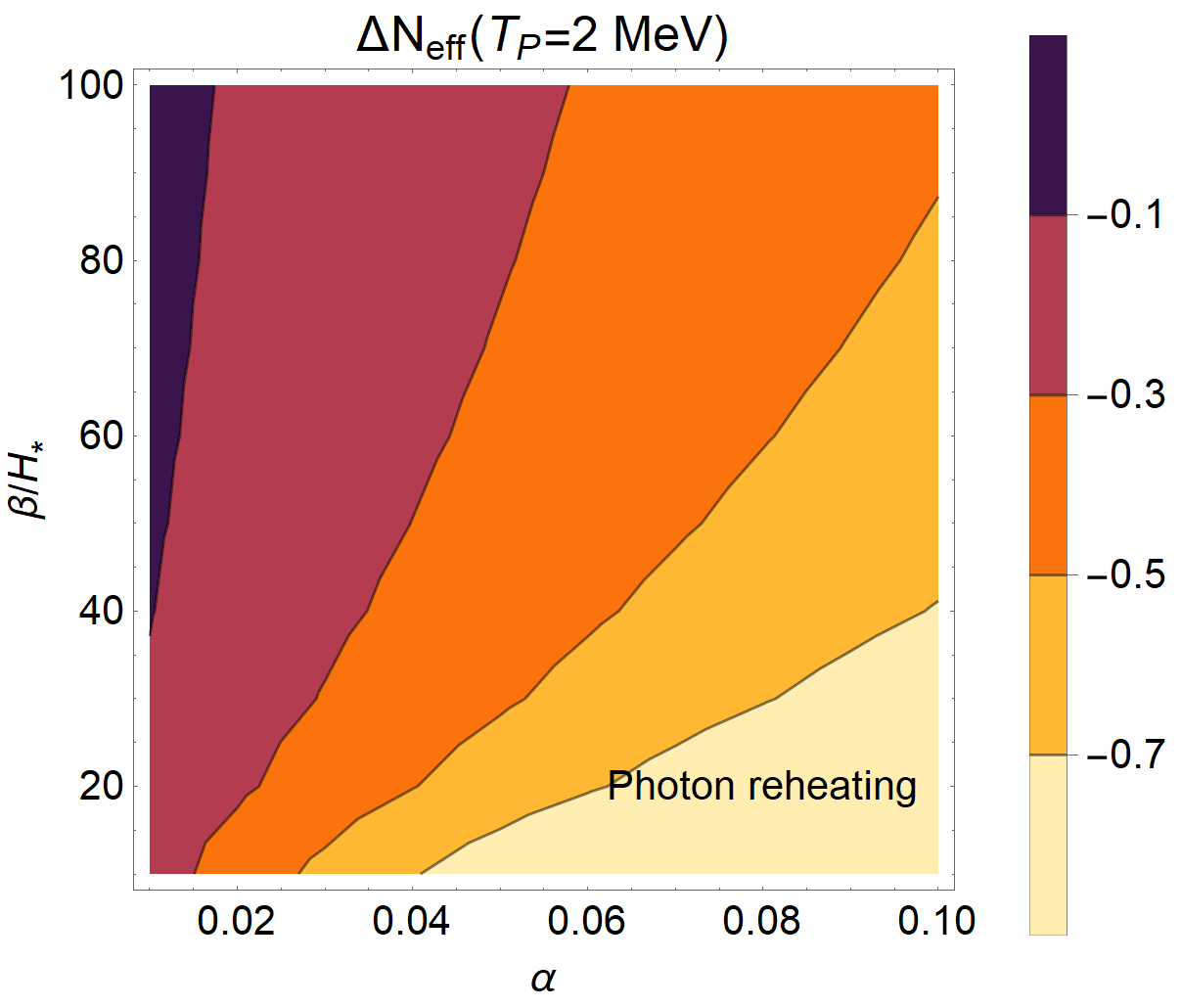}
\includegraphics[width=0.4\textwidth]{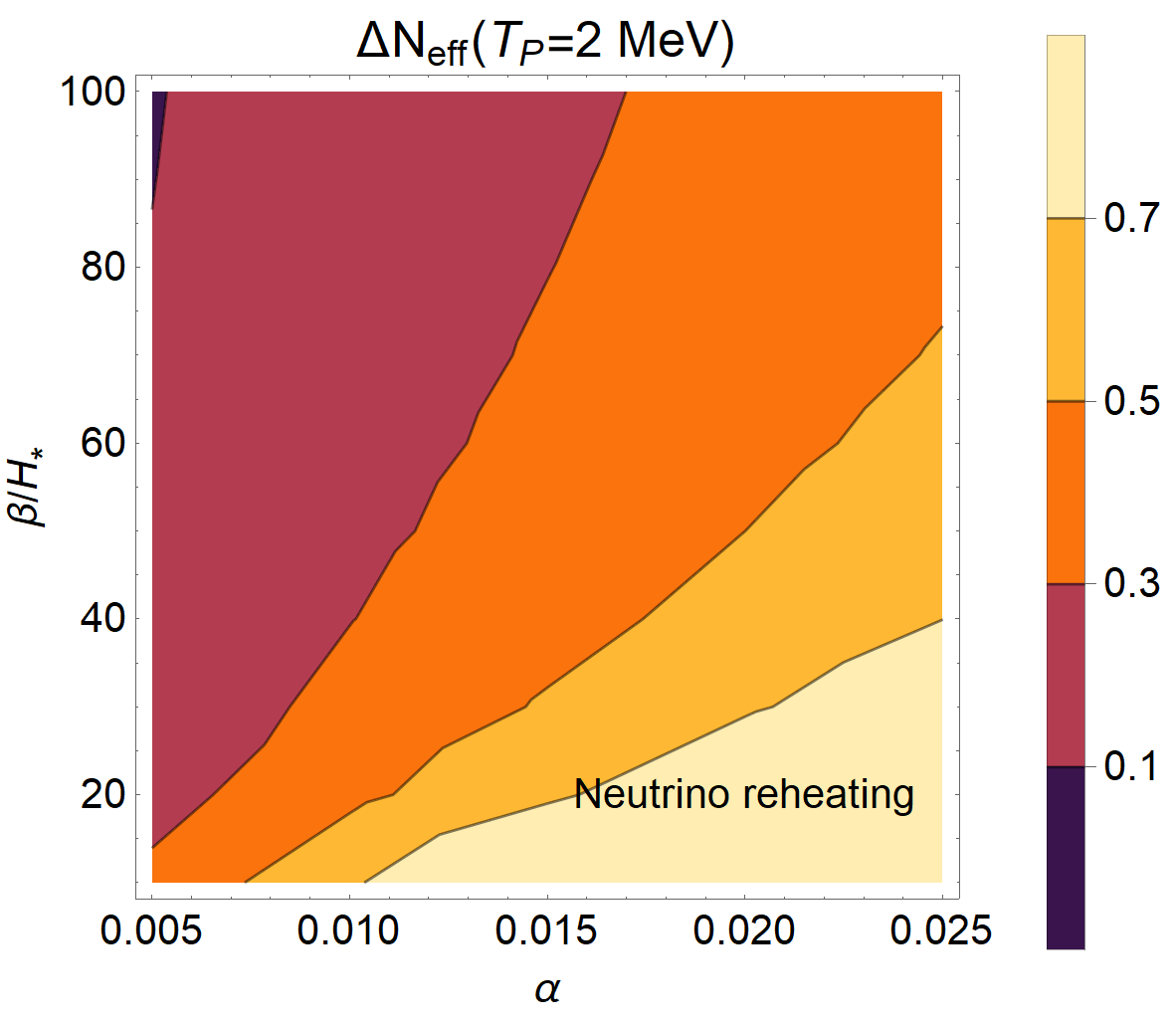}
		\caption{The panels show the value of $\Delta N_{\rm eff}=N_{\rm eff}-N^{\rm SM}_{\rm eff}$ (with $N^{\rm SM}_{\rm eff}=3.044(1)$~\cite{EscuderoAbenza:2020cmq,Akita:2020szl,Froustey:2020mcq,Bennett:2020zkv}) for photon reheating (left two plots) and neutrino reheating cases (right two plots) with $T_{\rm P}=1\,\text{MeV}$ and $T_{\rm P}=2\,\text{MeV}$. }
		\label{fig:DeltaNeff}
		\end{center}
	\end{figure*}

Fig.~\ref{fig:TnuT} illustrates the evolution of $T_\nu/T_\gamma$ with $T_\gamma$ for photon and neutrino reheating. Before neutrino decoupling, both neutrino and photon temperatures evolve similarly, resulting in $T_\nu=T_\gamma$ until neutrinos decouple and electron-positron annihilation. After the completion of electron-positron annihilation, the ratio $T_\nu/T_\gamma$ becomes constant (not equal to 1), and this temperature ratio determines the value of $N_{\rm eff}$. As depicted in the two plots, first-order PTs occurring around the MeV scales have a significant impact on neutrino decoupling during both photon and neutrino reheating. In the case of photon reheating, the injection of vacuum energy into photons leads to an increase in photon temperature, resulting in a decrease in the ratio of neutrino-to-photon temperatures ($T_\nu/T_\gamma$) and a decrease in $N_{\rm eff}$. The opposite effect is observed for neutrino reheating. 
 For Weakly Interacting Massive Particle (WIMP) with a mass above $20\,\text{MeV}$, the decoupling of neutrinos is almost unaffected since such a particle would decouple from the primordial plasma while it is non-relativistic, at temperatures around $T \sim m/20$~\cite{Kolb:1990vq}. 
The impact of dark matter with masses below $20\,\text{MeV}$ on the background thermodynamics is significant, and for the sake of clarity, we choose a vector boson with $m_\chi=1\,\text{MeV}$ for illustration. Generally, the appearance of dark matter would decrease (increase) the $N_{\rm eff}$ for the electrophilic (neutrinophilic) scenario~\cite{Sabti:2019mhn}. 
Therefore, both PTs and dark matter contribute to the injection of energy, which affects the process of neutrino decoupling. However, there is a noticeable discrepancy between the two effects due to the short duration of MeV-scale PTs compared to the evolution of dark matter. Consequently, the variation in the temperature ratio is sharper when PTs are present, as depicted in Fig.~\ref{fig:TnuT}, compared to the scenario without phase transitions but including $m_\chi=1\,\text{MeV}$ dark matter. In the following, we present the detailed effects of PT parameters on the deviation of $N_{\rm eff}$ from the Standard Model prediction when light dark matters' effect is negligible.
 

First-order PTs occurring at MeV scales could cause photon reheating or neutrino reheating.
To demonstrate the PT's effect on the neutrino decoupling process when $T_{\rm P}$ is close to $1\,\text{MeV}$, we present the cases of photon reheating and neutrino reheating by taking the vector dark matter $m_\chi=30$ MeV where the light degree freedoms' effect is negligible.
In the left (right) two plots of Fig.~\ref{fig:DeltaNeff}, we demonstrate the impacts of PT parameters $\alpha$ and $\beta/H_{*}$ on $N_{\rm eff}$ for photon (neutrino) reheating with different $T_{\mathrm{p}}$. The effects of the PT are stronger for slower PTs with lower $\beta/H_\star$ and stronger PTs with larger $\alpha$, and the effect of the PT is greater for the scenario of $T_{\mathrm{p}}=1\,\text{MeV}$ than that of $T_{\mathrm{p}}=2\,\text{MeV}$. In the case of photon reheating, $N_{\rm eff}$ would decrease due to the increase in photon temperature, and $N_{\rm eff}$ would increase for neutrino reheating.

\section{Primordial Nucleosynthesis and PTs}
The neutron fraction $X_n \equiv n_n/n_b$, which represents the ratio of neutron number density to baryon number density, is a crucial intermediate quantity in BBN. At high temperatures, $n\leftrightarrow p$ reactions are in equilibrium. Neglecting the chemical potential of electrons and neutrinos, the equilibrium abundance of neutrons is $X_n^{\rm eq}\approx e^{-Q/T}/(1+e^{-Q/T})$, where $Q\equiv m_n-m_p=1.293\,\text{MeV}$. After weak reactions freeze out at $T_{\rm FO}$ ($t_{\rm FO}$), $X_n$ gradually decreases due to occasional weak reactions and is eventually dominated by free neutron decay. During this period, the remaining fraction of neutrons is given by $X_{n}(t>t_{\rm FO})\approx X_{n}(t_{\rm FO})e^{-(t-t_{\rm FO})/\tau_{n}}$. As the temperature falls to $T_{\rm nuc}\approx 0.078\,\text{MeV}$ (at time $t_{\rm nuc}$), the abundance of deuterium reaches its maximum value, and the helium abundance begins to increase rapidly.

The appearance of MeV scale light degree of freedoms would change the energy density of the Universe and yield an underproduction of the mass density fraction of $^4H_e$ ($Y_{\rm P}$) and deuterium abundance (${\rm D/H}|_{\rm P}$) for the case of electrophilic dark matter. Meanwhile, for the case of neutrinophilic dark matter, one has an overproduction of the $Y_{\rm P}$ and ${\rm D/H}|_{\rm P}$ (see, e.g., Refs.~\cite{Kolb:1986nf,Serpico:2004nm,Nollett:2013pwa,Nollett:2014lwa,Sabti:2019mhn}). In the neutrino sector, there could be a significant electron-neutrino asymmetry which can directly impact the neutron-to-proton ratio~\cite{Kohri:1996ke,Simha:2008mt,Fields:2019pfx}, given by $n_n/n_p\approx \exp(-Q/T-\xi_{\nu_e})$, where $\xi_{\nu_e}\equiv \mu_{\nu_e}/T_{\nu_e}$ represents the ``degeneracy parameter" of the neutrino chemical potential. Here, we neglect the chemical potential 
of electrons which is highly constrained considering the electric charge neutrality of the early Universe. It is evident that a positive (negative) $\xi_\nu$ decreases (increases) the ratio, and consequently reduces (enhances) the final abundances of $Y_{\rm P}$ and ${\rm D/H}|_{\rm P}$. 
In the following, we analyze the effects of first-order PTs on primordial nucleosynthesis, since we have additional energy injection during the PTs, which changes the time-temperature relation around the MeV scale.

 \begin{figure*}[!htp]
\begin{center}
\includegraphics[width=0.4\textwidth]{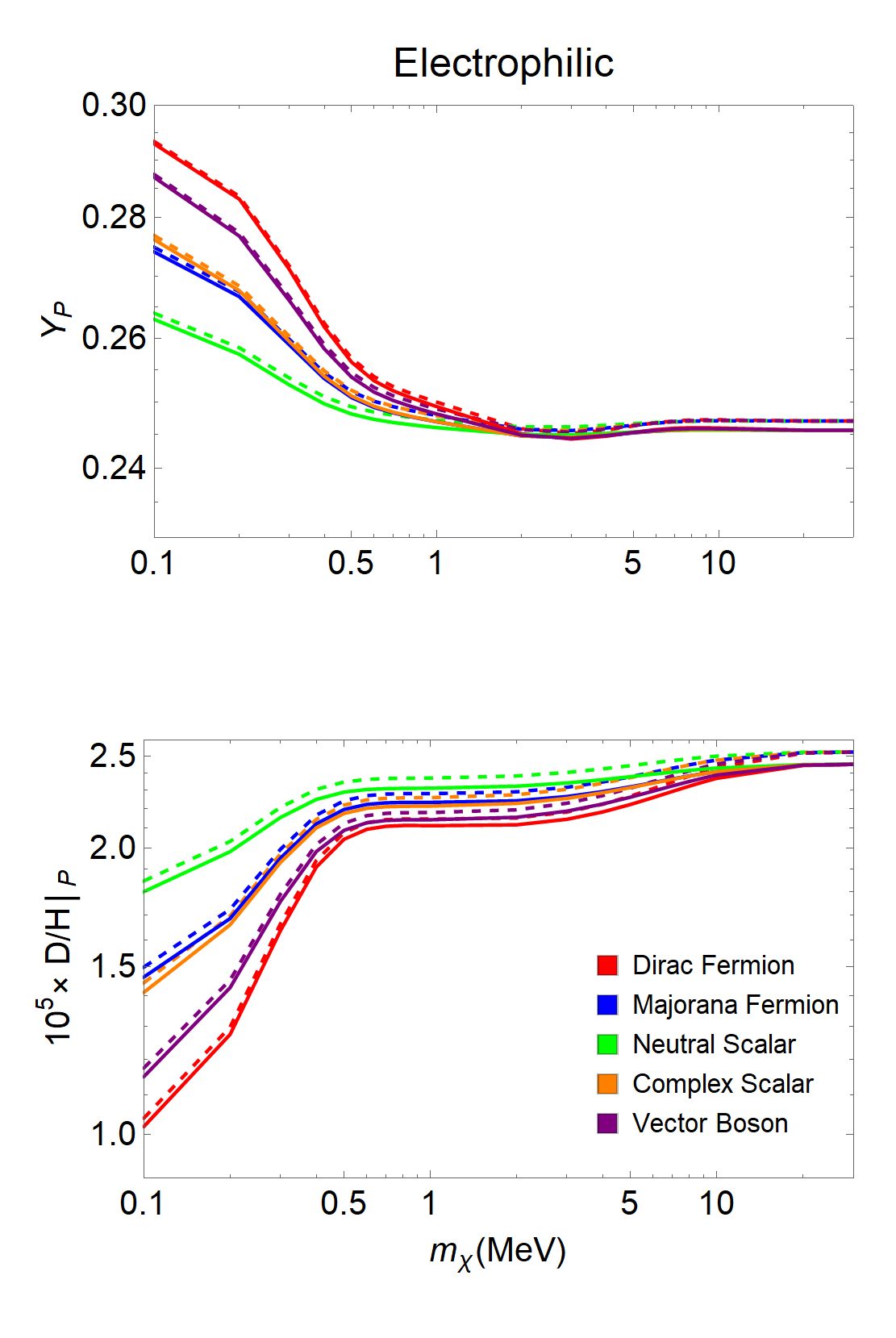}
\includegraphics[width=0.4\textwidth]{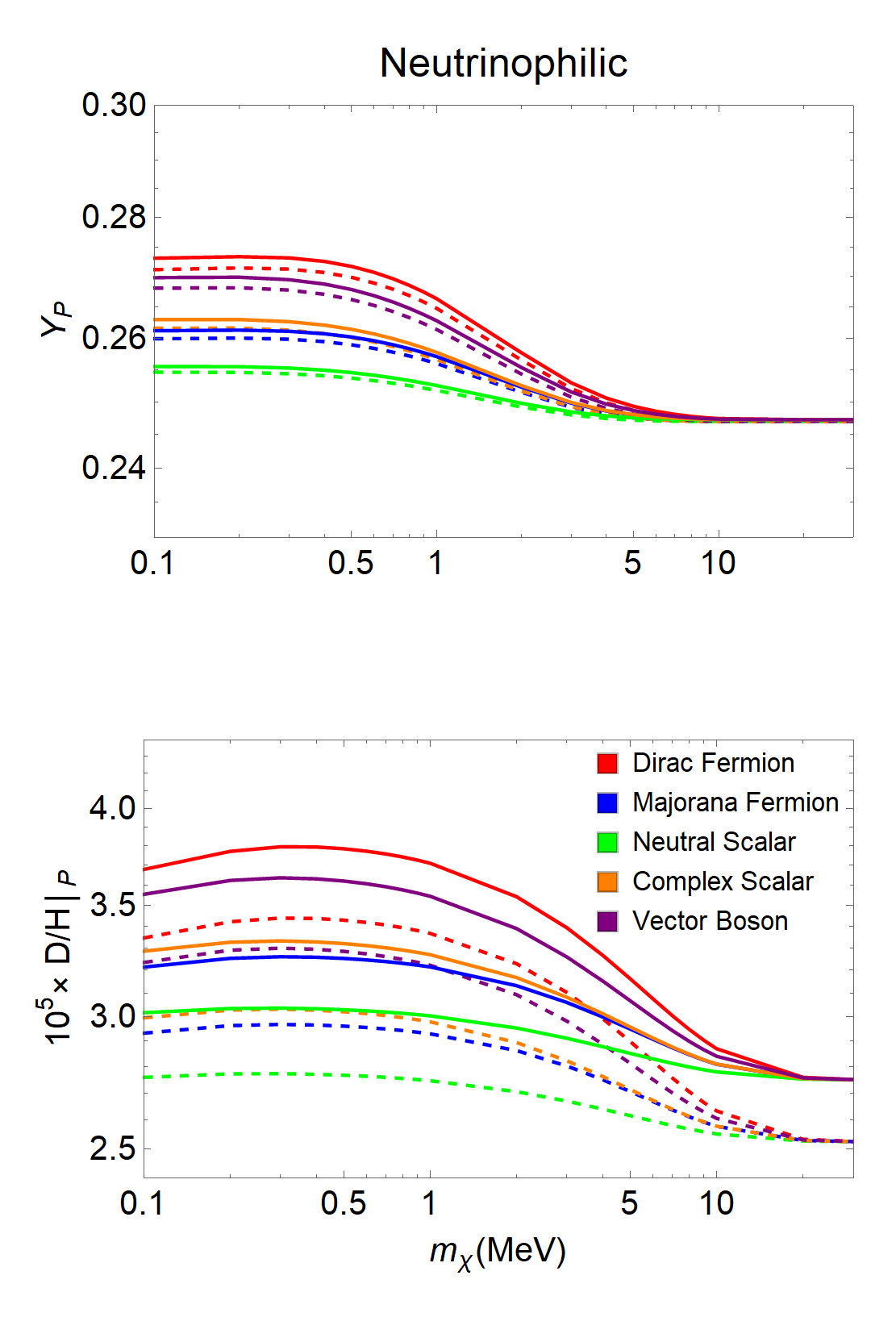}
		\caption{Impacts of light BSM particles on primordial nucleosynthesis as a function of their mass $m_{\chi}$. The \textit{top  (bottom) two panels} correspond to $Y_{\rm P}$ (${\rm D/H}|_{\rm P}$). The dashed line corresponds to the case where no PT is considered, the solid line corresponds to the scenarios with a PT where $T_{\rm P} = 1\,\text{MeV}$,$\alpha = 0.01$, and $\beta/H_{*} = 10$ for both photon reheating (left) and neutrino reheating (right).}
		\label{fig:BBNcomparison}
		\end{center}
		\end{figure*}
		
In the case of photon reheating from first-order PTs, the total energy density and the Hubble rate decrease for a given photon temperature $T=T_{\gamma}$, so is the magnitude of $dT/dt$. Therefore the nucleosynthesis will occur later than in the Standard Model, i.e., $t_{\rm nuc}$ is delayed, and more neutrons decayed, resulting in a smaller $Y_{\rm P}$. 
 On the other hand, the neutrino density is lower than the scenario without PTs, resulting in a decrease in the weak reaction rate $\Gamma_{np}$ at a given T, which can lead to an earlier freeze-out, with larger neutron fraction $X_n(T_{\rm FO})$. Though the reduced neutrino density also leads to a decrease in the Hubble rate which partially offsets the effect on $\Gamma_{np}$~\cite{Ichikawa:2005vw}, the reduction effect of the weak rate $\Gamma_{np}$ is more significant due to its high dependence on temperature~\cite{Huang:2021dba}. Considering these two points together, the freeze-out temperature $T_{\rm FO}$ would be higher than the scenario in the absence of PTs, i.e., freeze-out occurs earlier, which probably yields a larger final $Y_{\rm P}$.
 For neutrino reheating case from first-order PTs, photon temperature changes faster with time, so $t_{\rm nuc}$ is advanced, which increases final $Y_{\rm P}$, and the $T_{\rm FO}$ would be lower than the case without PTs. 
In summary, taking into account the effects of PTs, variations in both $X_{n}(t_{\rm FO})$ and $t_{\rm nuc}$ lead to changes in the value of $X_n(t_{\rm nuc})$, thereby influencing the ultimate helium abundance which is approximately given by $Y_{\rm P}\approx 2X_n(t_{\rm nuc})$.
The peak of deuterium abundance occurs at around $T_{\rm nuc}$ which corresponds to a later (earlier) $t_{\rm nuc}$ for photon (neutrino) reheating. After that, deuterium abundance starts to decrease. The primary factor that affects the final abundance of deuterium is the time of the destruction of deuterium, the reheating of photons (neutrinos) triggered by the first-order PTs will affect the time-temperature relation, causing a more (less) time to destroy deuterium, which leads to a smaller (larger) final ${\rm D/H}|_{\rm P}$.

For the calculation of the primordial nucleosynthesis, we pass the necessary thermodynamic parameters including $T_{\gamma}$, $T_{\nu}$, the scale factor $a$, and the Hubble parameter $H$ obtained with the modified  \texttt{{nudec}\_BSM} on to the BBN code \texttt{PRIMAT}~\cite{Pitrou:2018cgg}. These parameters were constructed as a function of time using the interpolation method and were used to replace the original thermodynamics of \texttt{PRIMAT}. By doing so, the time evolution of the 
nuclei abundances were calculated by recomputing weak interactions and nuclear reaction rates.

We verified the correctness of our modified version by generating curves for the primordial helium and deuterium abundances with dark matter mass, which is in agreement with the results presented in Fig.1 of Ref.~\cite{Sabti:2019mhn} when the PT dynamics were not considered. 
In Fig.\ref{fig:BBNcomparison}, we show specifically the effect of photon reheating  and neutrino reheating first-order PT on BBN. Where, the $Y_{\rm P}$ and ${\rm D/H}|_{\rm P}$ predictions are calculated with $\Omega_\mathrm{b} h^2 = 0.021875$ and $\tau_n = 879.5\,\text{s}$~\cite{Sabti:2019mhn}. Compared to the scenarios in the absence of PT, we find both the magnitudes of $Y_{\rm P}$ and ${\rm D/H}|_{\rm P}$ decrease for photon reheating, while the opposite is true for neutrino reheating.
 Though the effect of PT on ${\rm D/H}|_{\rm P}$ is found to be consistent with the results in Ref.~\cite{Bai:2021ibt}, the effect of photon reheating on $Y_{\rm P}$ is different, which is related to the two parameters mentioned before, $X_{n}(t_{\rm FO})$ and $t_{\rm nuc}$. In our case, the effect of $t_{\rm nuc}$ is a little stronger than $X_{n}(t_{\rm FO})$ and leads to a decrease in $Y_{\rm P}$, however, in the case of Ref.~\cite{Bai:2021ibt}, the effect of $X_{n}(t_{\rm FO})$ is stronger than $t_{\rm nuc}$ and finally leads to a larger $Y_{\rm P}$. We further note that these plots confirm that the effects of different dark matter species are negligible for large mass since all the predictions of $Y_{\rm P}$ and ${\rm D/H}|_{\rm P}$ merges at $m_\chi\gtrsim 20$ MeV for both the scenarios with and without considering PTs.

\section{Constraints on PT parameters}

\begin{figure*}[!htp]
\begin{center}
\includegraphics[width=0.4\textwidth]{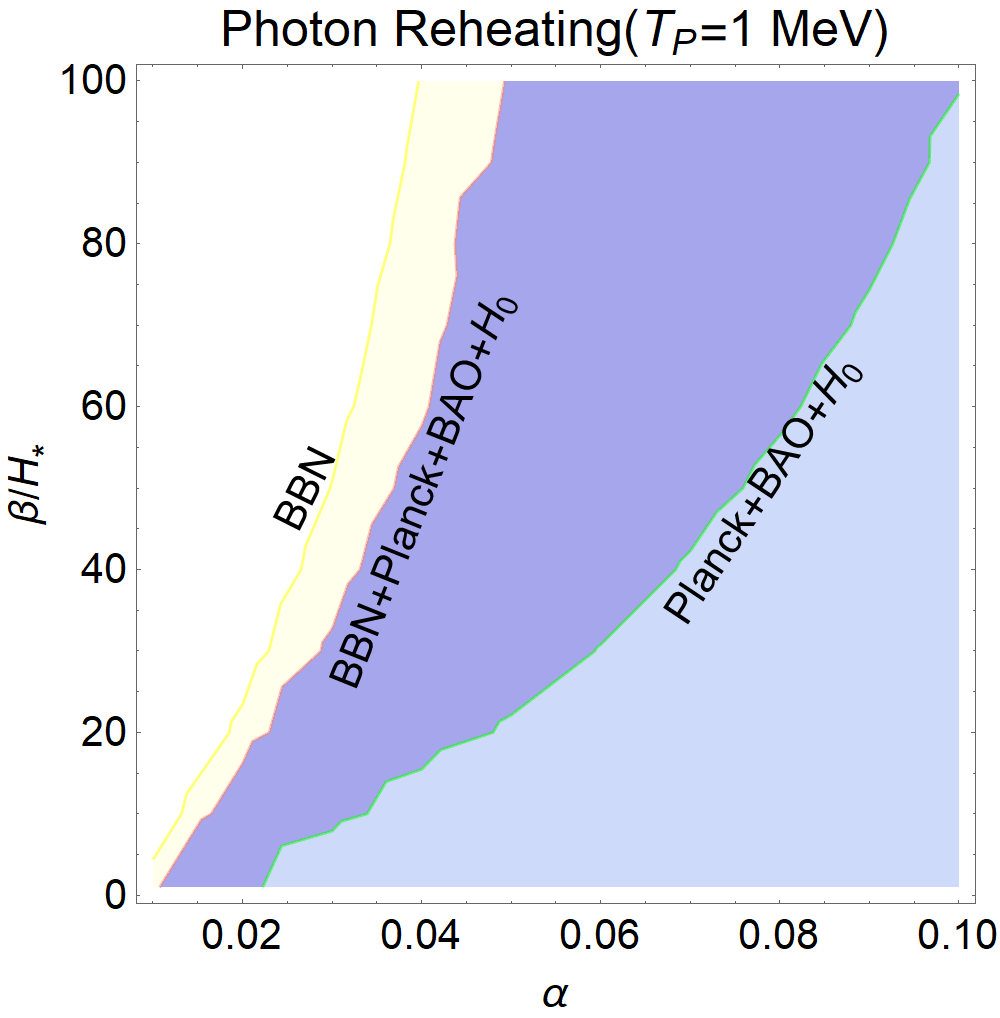}
\includegraphics[width=0.4\textwidth]{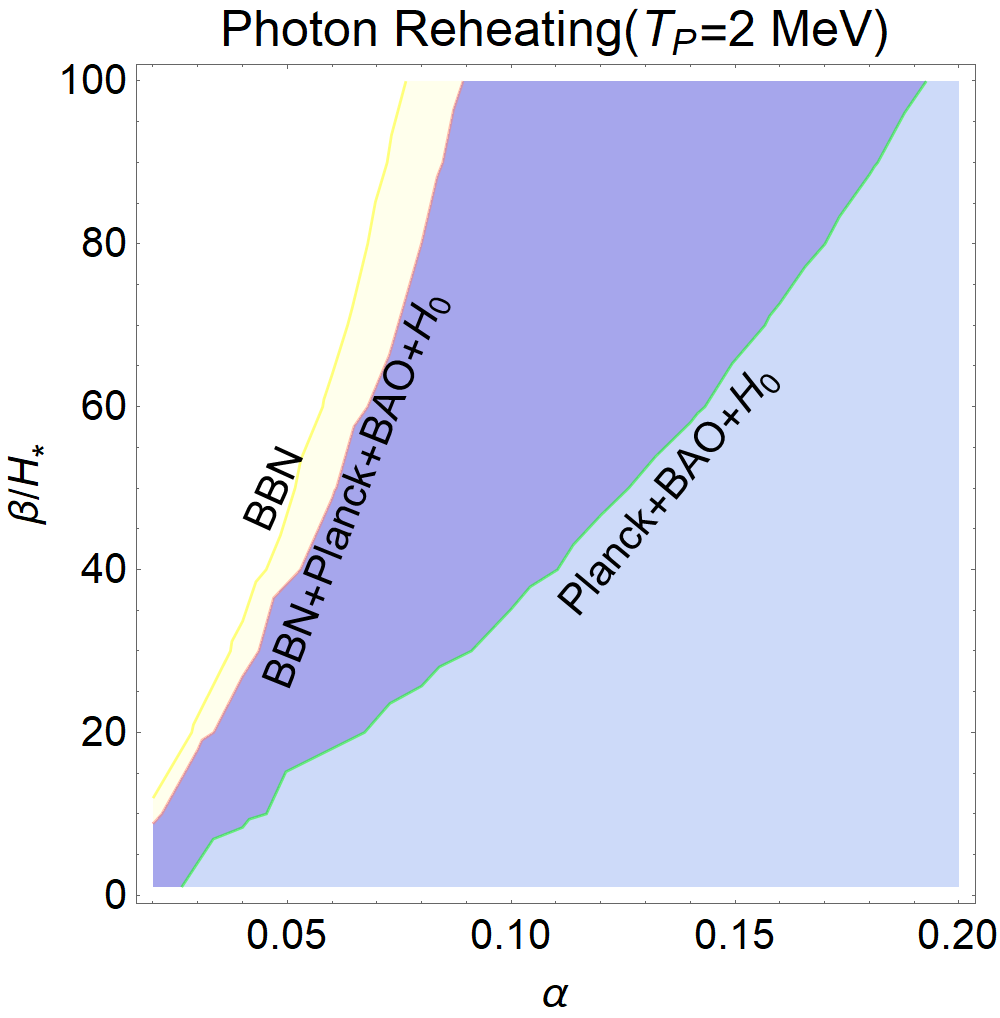}
\includegraphics[width=0.4\textwidth]{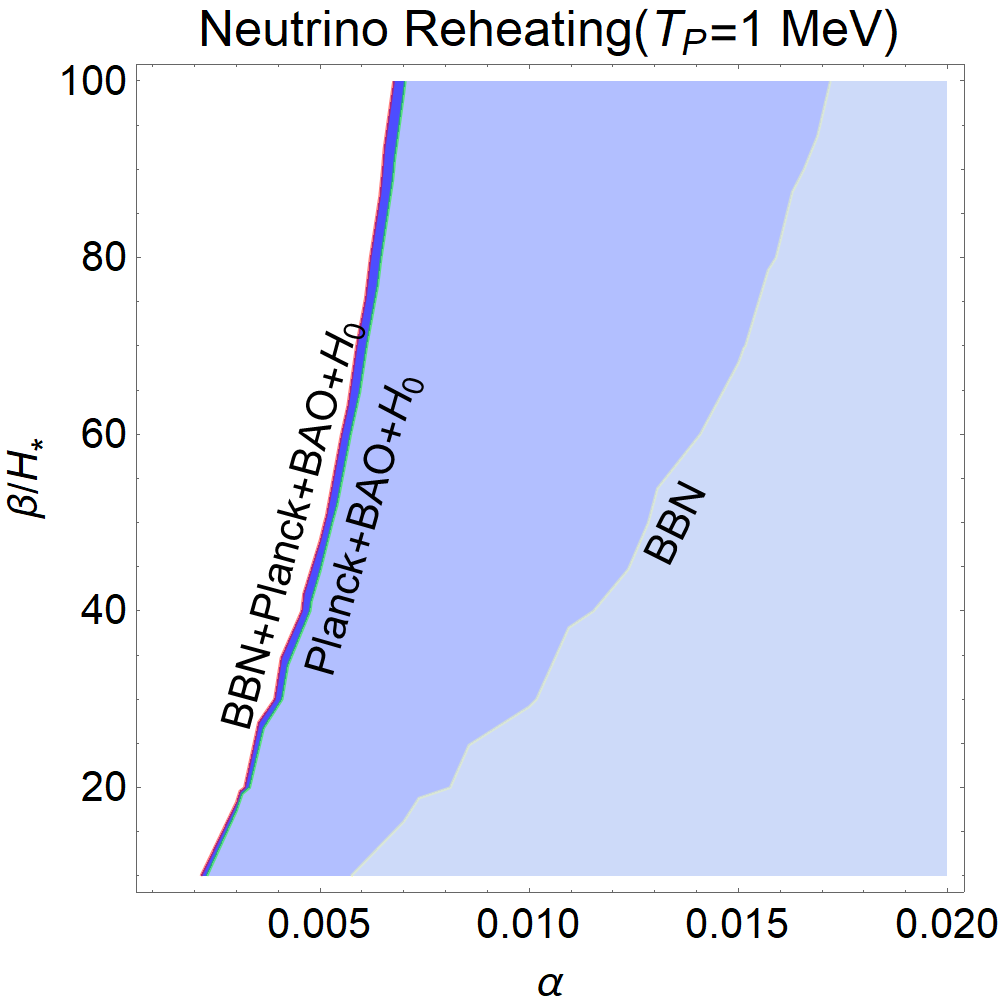}
\includegraphics[width=0.4\textwidth]{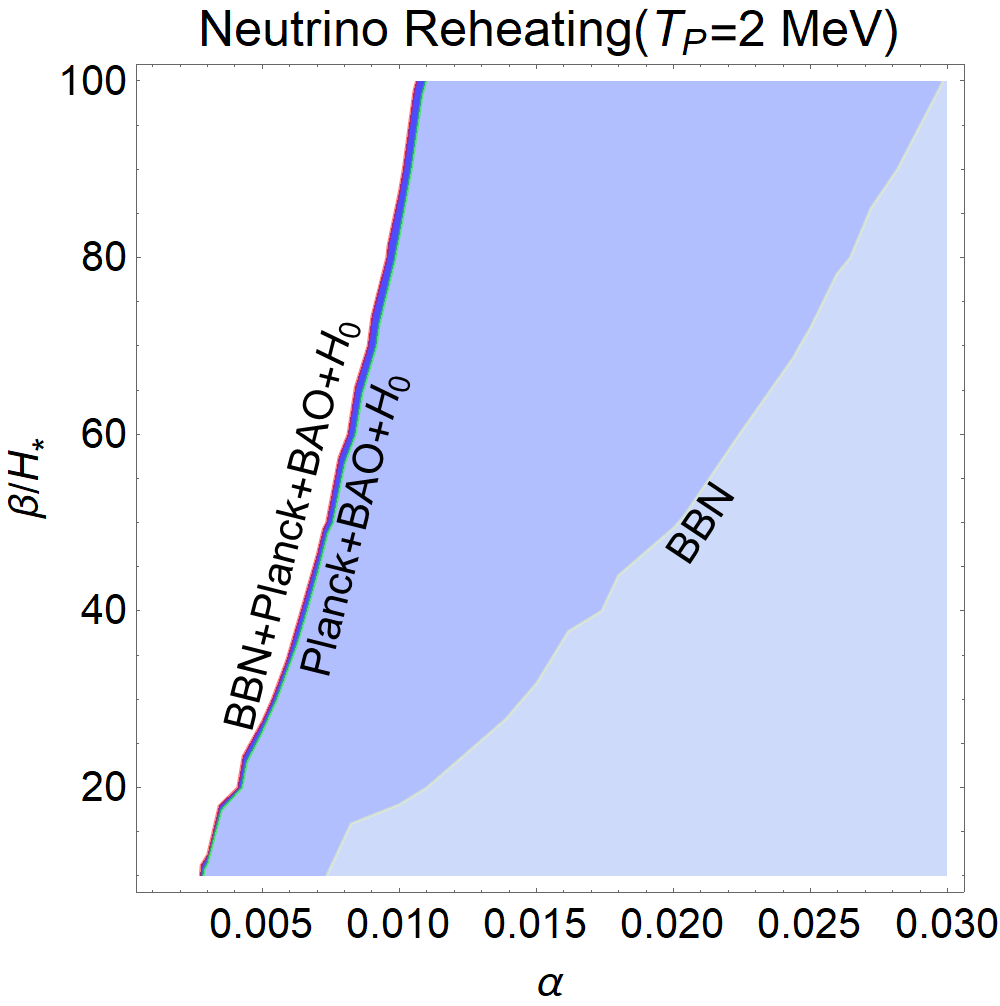}
		\caption{The $95\%$ CL constraints on the PT parameters $\alpha$ and $\beta/H_{*}$ from CMB and BBN datasets at $T_{\rm P}=1\,\text{MeV}$ (left) and $T_{\rm P}=2\,\text{MeV}$ (right) for both photon and neutrino reheating case. The three exclusion lines correspond to BBN (yellow), Planck+BAO+$H_0$ (green), and BBN+ Planck+BAO+$H_0$ (red), respectively, which exclude the area on the right.}
		\label{fig:BBNCMBX2}
		\end{center}
	\end{figure*}
	
	In this section, we study constraints on the low-scale first-order PTs, light dark matter, and the lepton asymmetry from the BBN and CMB observations.
For the analysis of the BBN, we consider the observation of primordial abundances of helium and deuterium $(Y_{\rm P},{\rm D/H}|_{\rm P})$.    
And the CMB observations precisely measure the value of $(\Omega_{b}h^{2}, N_{\text{eff}}, Y_{\rm P})$. 
The local measurement of the $H_0$ from the SH0ES collaboration~\cite{Riess:2019cxk} would uplift the reconstructed value of the effective neutrino number for some amount, for the neutrino interpretation of the Hubble Tension we refer to Ref.~\cite{Blinov:2019gcj}. We consider the constraints from BBN, CMB, and the joint constraints from BBN+CMB which are simply constructed by the sum of $\chi^2_{\text{BBN}}$ and $\chi^2_{\text{CMB}}$. For details see Appendix.

We first solely investigate the dynamics and effects of phase transitions by taking vector dark matter with the fixed mass of $m_\chi=30\,\text{MeV}$, where the light degree freedoms effect is negligible. 		
 Fig.~\ref{fig:BBNCMBX2} displays the exclusion limits at the 95\% confidence level (CL) for $\alpha$ and $\beta/H_*$ after marginalizing over $\Omega_b h^2$. By setting $\Omega_b h^2=0.021875$, we obtain the minimum $\chi^2$ value, denoted as $\chi^2_{\rm min}$, and the corresponding 95\% CL limits are defined by $\Delta \chi^{2}= \chi^{2} -\chi^{2}_{\rm min} = 5.99$. 
 We find that: 1) strong PT of relatively large $\alpha$ and slow PT with small $\beta/H_\star$ are excluded; 2) BBN and CMB observations yield weaker constraints on PTs occurring at higher temperatures (with larger $T_{\rm P}$);
 and 3) the constraint from BBN is stronger than that of CMB for photon reheating, this is because that ${\rm D/H}|_{\rm P}$ provides stronger constraints compared to $N_{\rm eff}$, while CMB provides stronger constraint than BBN for neutrino reheating which is caused by that neutrino reheating leads to a larger change on $N_{\rm eff}$ than photon reheating for the same energy injection. And this also explains why the allowed PTs parameter space for neutrino reheating is smaller much than photon reheating.
For the photon reheating PTs with $\beta/{H_*}=50$, the BBN data set constrain the PT strength to be $\alpha \lesssim 0.035$ at $T_{\rm P} = 1\,\text{MeV}$ and  $\alpha \lesssim  0.06$ at $T_{\rm P} = 2\,\text{MeV}$.


	\begin{figure}[!htp]
\begin{center}
\includegraphics[width=0.4\textwidth]{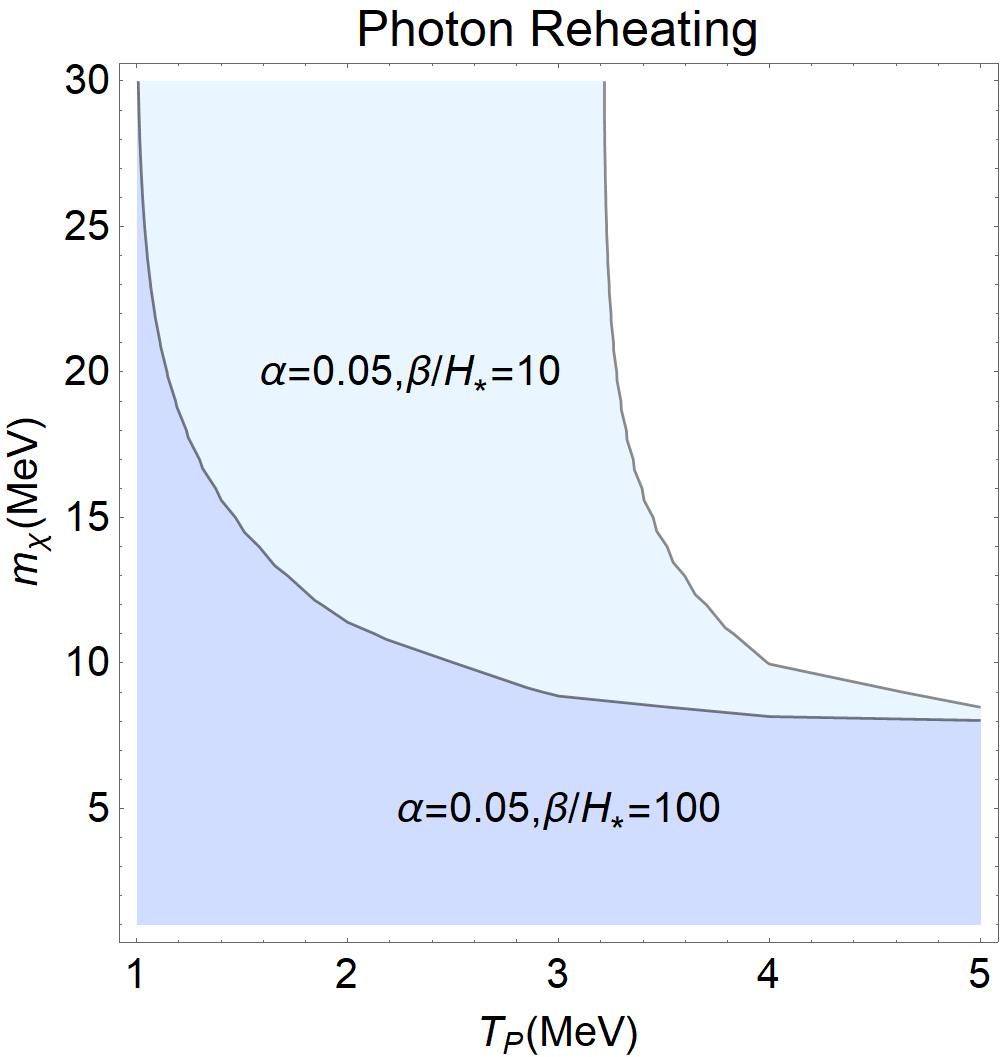}
\includegraphics[width=0.4\textwidth]{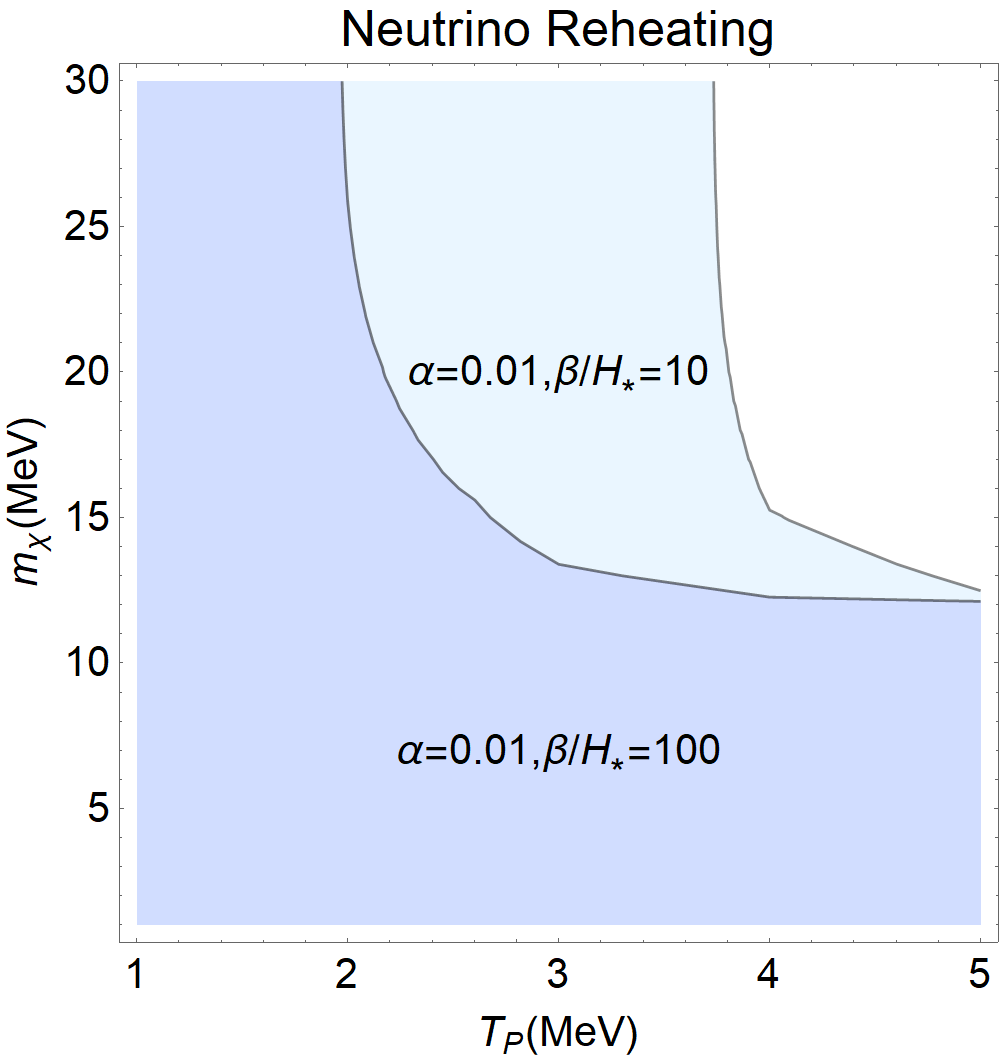}
		\caption{The shading regions show the $95\%$ CL exclusion on vector boson mass and PT temperature $T_{\rm P}$ from combining BBN+CMB data by constructing a joint $\chi^2$. The top (bottom) plot corresponds to photon (neutrino) reheating driven by PTs. For the photon (neutrino) reheating case, we fix the PT strength $\alpha=0.05$ ($\alpha=0.01$), with $\beta/H_\star=10$ or $100$.}
		\label{fig:TPMX2}
		\end{center}
	\end{figure}

 \begin{figure}[!htp]
\begin{center}
\includegraphics[width=0.43\textwidth]{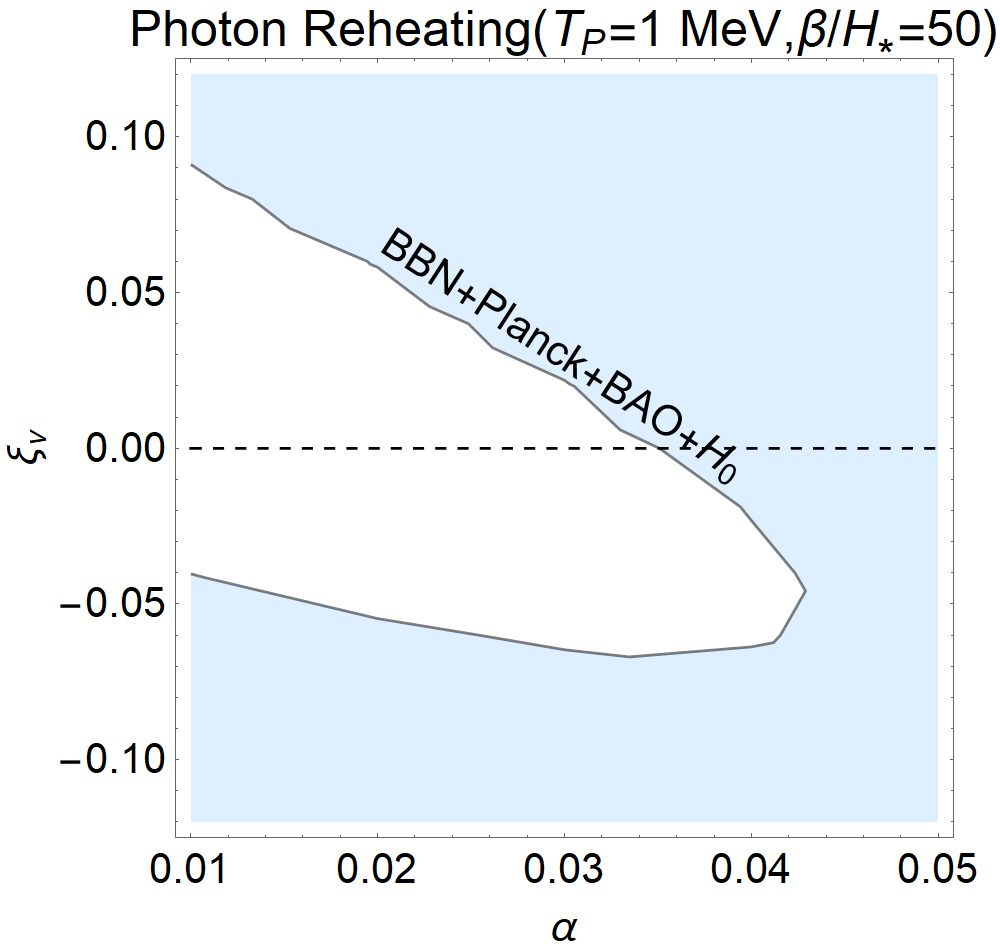}
\includegraphics[width=0.423\textwidth]{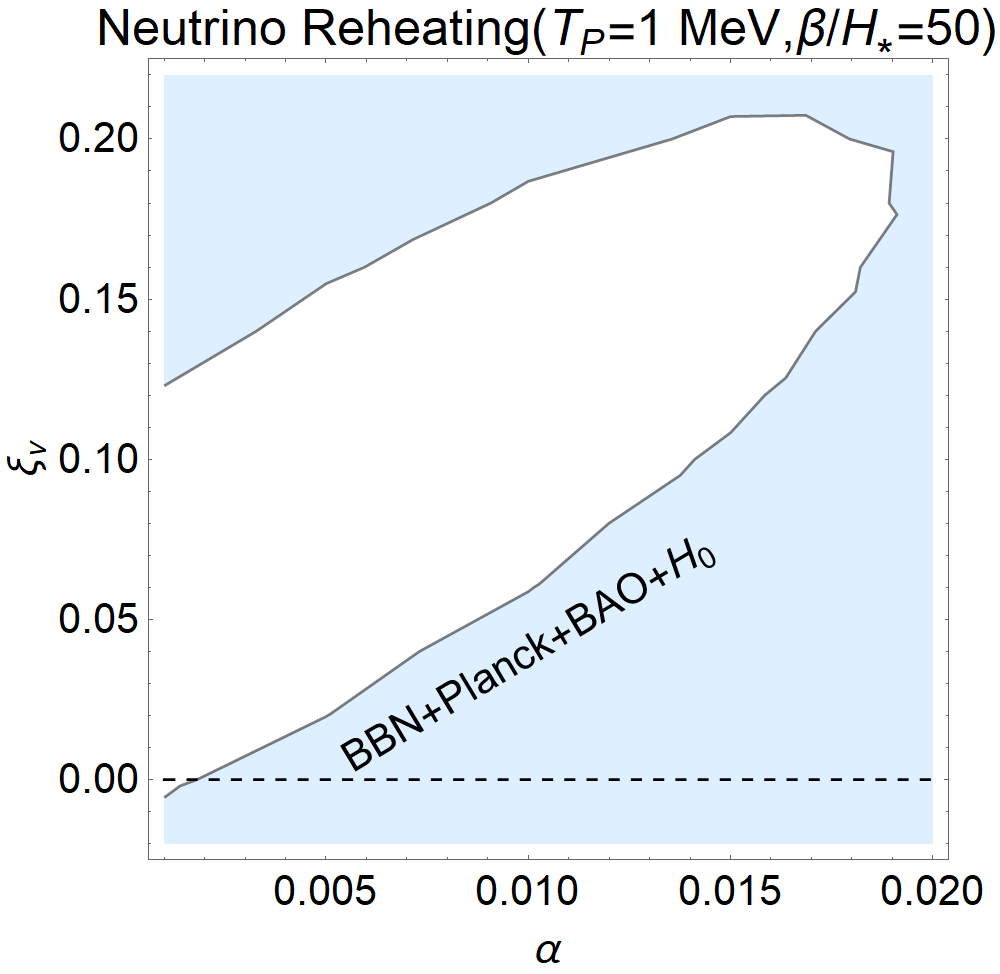}
		\caption{The joint constraints on PTs and lepton asymmetry with $T_{\rm P}=1\,\text{MeV}$ and $\beta/H_\star=50$ considering combining BBN+CMB datasets. The dashed lines mean $\xi_\nu=0$, that is just the case of Fig.~\ref{fig:BBNCMBX2} with the same $T_{\rm P}$ and $\beta/H_\star$.}
		\label{fig:AXIX2}
		\end{center}
	\end{figure}

To consider the synergy effect from the thermal light dark matter and PTs, we study the effects of dark matter and PTs simultaneously. For the study, we consider PTs originating from the dark sector, where photon (neutrino) reheating is induced by electrophilic (neutrinophilic) vector dark matter.  
Fig.~\ref{fig:TPMX2} presents the $95\%$ confidence level constraints from the combination of BBN and CMB datasets on the PTs temperature $T_{\rm P}$ and the dark matter mass $m_\chi$. In the case of slow PTs with small $\beta/H_\star$, the constraints on $m_\chi$ are more stringent. As $T_{\rm P}$ increases, the exclusion curves become more horizontal. For $T_{\rm P}\gtrsim 4\,\text{MeV}$, the impact of the PTs becomes weak. In these cases, the allowed range of dark matter mass is $\gtrsim 8(12)\,\text{MeV}$ for photon (neutrino) reheating.


In Fig.~\ref{fig:AXIX2}, we analyze the combined effects of the lepton asymmetry from the neutrino sector and the PTs. For illustration, we choose vector dark matter with $30\,\text{MeV}$, and fix PT temperature ($T_{\rm P}=1\,\text{MeV}$) and PT duration ($\beta/H_\star=50$) with a free PT strength $\alpha$. We present the constraints on PT strength $\alpha$ and the degeneracy parameter $\xi_\nu$ (we drop the subscript ``e" in the $\xi_{\nu_e}$) from the BBN+CMB datasets. Here, we neglect the impact of the lepton asymmetry on $N_{\rm eff}$ since its modification is relatively small~\cite{Escudero:2022okz,Burns:2022hkq}, and the modification on $N_{\rm eff}$ is primarily caused by the PTs' reheating effect.
As stated before and found by Ref.~\cite{Pitrou:2018cgg}, a positive (negative) lepton asymmetry can lead to a relatively small (large) value of $Y_{\rm P}$ and the ${\rm D/H}|_{\rm P}$. In comparison with the case where only PT's effect is considered (as shown in Fig.~\ref{fig:BBNCMBX2}), we observe that a positive (negative) $\xi_\nu$ roughly tightens (loosens) the constraints on PTs in the case of photon reheating, whereas the opposite holds for neutrino reheating.
For instance, in the case of photon reheating with $\xi_\nu=0.05$, the PT strength $\alpha$ is constrained to be $\leq$0.022, while in the case of neutrino reheating with $\xi_\nu=0.15$, $\alpha$ fall within the range of [0.004,0.018]. 
And, with the increase of the PT strength $\alpha$, one has more stringent constraints on $\xi_\nu$. Explicitly, the allowed regions of the degeneracy parameter are $\xi_\nu\sim[-0.06,0.1]$ for $\alpha\sim[0.01,0.04]$ in the photon reheating case and $\xi_\nu\sim[0,0.2]$ for $\alpha~\sim[0,0.02]$ in the neutrino reheating scenario.

\section{ Conclusion and discussion}
In this work, we observe that: 1) the thermal dynamics of first-order PTs lead to changes in the BBN and CMB predictions; 2) the constraints on the thermal light dark matter and the neutrino lepton asymmetry are more rigorous when the effects of the PTs are included. More explicitly, in comparison with the CMB observations, the BBN observations are stronger (and weaker) for the photon (and neutrino) reheating scenarios. The CMB and BBN constrain the PT strength to be around $\sim \mathcal{O}(10^{-2})$ (and $\mathcal{O}(10^{-3}$)) for photon (and neutrino) reheating scenarios at MeV scale. The slow PT with low $\beta/H_\star$ and the low PT temperature cases suffer stringent limits from the BBN and CMB observations. The appearance of slower first-order PT with lower PT temperature would yield a stricter limit on the dark matter of larger mass. The magnitude of the neutrino lepton asymmetry is limited to a much smaller range roughly in the range of $[-0.06,0.1]$ (and  [0,0.2]) for photon (and neutrino) reheating scenario depending on the PT strength.

In the future, some proposed ground-based CMB experiments, such as the Simons Observatory~\cite{SimonsObservatory:2018koc} and CMB-S4~\cite{CMB-S4:2016ple,Abazajian:2019eic}, would provide a more precise determination of $\Omega_b h^2$, $N_{\rm eff}$, and $Y_{\rm P}$, which might set stronger constraints on PTs, thermal light dark matter, and the lepton asymmetry.
The observations of the curvature perturbation at CMB would yield constraints on low-scale dark sectors considering the super-horizon effects of the slow PTs
~\cite{Liu:2022lvz,Ramberg:2022irf}.  
Nanohertz gravitational wave detection conducted by PPTA, NANOGrav, and SKA would have the chance to probe the low-scale PTs. Refs.~\cite{NANOGrav:2021flc,Xue:2021gyq} provide the constraints on slow and strong first-order PTs occurring around the QCD scale. This study is complementary to these studies and provides much stronger constraints on strong and slow PTs occurring close to the MeV scale. It was noted that the free streaming of neutrinos would damp gravitational waves~\cite{Weinberg:2003ur}.  In comparison with the scenario of purely first-order PT without considering the neutrino decoupling effects, the low-frequency tail of the gravitational wave spectrum from MeV scale PTs under study would be modified~\cite{Loverde:2022wih,Hook:2020phx}, which could be probed by Pulsar timing arrays soon.

\noindent{\it \bfseries Acknowledgments.}
	We are grateful to Miguel Escudero Abenza and James Alvey for helpful discussions on the BBN code \texttt{PRIMAT} and the study of the neutrino decoupling with the code \texttt{nudec\_BSM}. We thank Shun Zhou for the insightful discussion on the relationship between neutrino decoupling and low-scale phase transitions. We thank Zach Weiner for bringing our attention to the relationship between neutrino-free steaming and gravitational waves.
	This work is supported in part by the National Key Research and Development Program of China Grants No. 2021YFC2203004, and in part by the National Natural Science Foundation of China under grants Nos. 12075041, 12147102, and 12322505.
	
	\section{Appendix}
To obtain the current constraints on low-scale PT from BBN observables, we take the effective BBN $\chi^2$ being~\cite{Sabti:2019mhn},~\cite{Sabti:2019mhn},
 \begin{equation}
\label{eq:chi2BBN}
\begin{aligned}
    \chi^{2}_{\rm BBN}&= \frac{\left[Y_{\rm P}(\Omega_bh^2, \alpha, \beta/H_{*}) - Y_{\rm P}^{\text{obs}}\right]^{2}}{ \sigma(Y_{\rm P}^{\text{th}})^{2}+\sigma(Y_{\rm P}^{\text{obs}})^{2}}\\
    &+ \frac{\left[\rm D/\rm H|_{\rm P}(\Omega_bh^2, \alpha, \beta/H_{*}) - \rm D/\rm H|_{\rm P}^{\text{obs}}\right]^{2}}{\sigma(\rm D/\rm H|_{\rm P}^{\text{th}})^{2}+\sigma(\rm D/\rm H|_{\rm P}^{\text{obs}})^{2}} \, .
\end{aligned}
\end{equation}
Where, the central values are: $Y_{\rm P}^{\rm obs} = 0.245 \,,\rm D/ \rm H|_{\rm P}^{\rm obs}=2.547\times 10^{-5}$, and the current observational errors are $\sigma(Y_{\rm P}^{\rm obs}) = 0.003, \sigma(\rm D/\rm H|_{\rm P}^{\text{obs}})= 0.025 \times 10^{-5}$~\cite{ParticleDataGroup:2020ssz},    
and theoretical errors are taken from Ref.~\cite{Pitrou:2020etk}: $\sigma(Y_{\rm P}^{\rm th}) = 0.00014,
\sigma(\rm D/\rm H|_{\rm P}^{\text{th}}) = 0.037 \times 10^{-5} $.	
To obtain the constraints on PT parameters from the CMB measurements of $(\Omega_{b}h^{2}, N_{\text{eff}}, Y_{\rm P})$, we take the Gaussian likelihood as~\cite{Sabti:2019mhn}\,,
\beqa
\label{eq:chi2CMB}
    \chi^{2}_{\text{CMB}} = (\Theta-\Theta_{\rm obs})^{\rm T} \,\Sigma_{\text{CMB}}^{-1}\, (\Theta-\Theta_{\rm obs}) \,,
\eeqa
with $\Theta \equiv (\Omega_{b}h^{2}, N_{\text{eff}}, Y_{\rm P})$  and 
\beqa
\quad\Sigma_{\text{CMB}} = \begin{bmatrix}
    \sigma_{1}^{2} & \sigma_{1}\sigma_{2}\rho_{12} & \sigma_{1}\sigma_{3}\rho_{13} \\
    \sigma_{1}\sigma_{2}\rho_{12} & \sigma_{2}^{2} & \sigma_{2}\sigma_{3}\rho_{23} \\
    \sigma_{1}\sigma_{3}\rho_{13} & \sigma_{2}\sigma_{3}\rho_{23} & \sigma_{3}^{2}
    \end{bmatrix} ~.
\eeqa
We take Planck+BAO+$H_0$ dataset with the experimental value of $\Theta$ being:
$\Theta_{\rm obs} =(0.02345, 3.36, 0.249)$, the parameters of the covariance matrices are $(\rho_{12},\rho_{13},\rho_{23}) = (0.011,0.50,-0.64)$ and $ (\sigma_{1},\sigma_{2},\sigma_{3})= (0.00025,0.25,0.020)$.

	\bibliography{ptbbn}

\end{document}